\documentclass[a4paper,12pt]{article}
\usepackage{geometry}
\usepackage[cp1251]{inputenc}
\usepackage{epsfig}
\usepackage{graphicx}

\begin{document}

\section*{
\begin{center}
Rotation Effects in Classical T Tauri Stars
\end{center}
}

\begin{center}
S. A. Artemenko, K. N. Grankin, P. P. Petrov
\end{center}

\begin{center}
\it{Crimean Astrophysical Observatory, Ukraine}
\end{center}

Surface temperature inhomogeneities in classical T Tauri stars (CTTS) induced by magnetic activity and mass accretion lead to rotational modulation of both photometric and spectroscopic parameters of these stars. Using the extended photometric catalogue by Grankin et al. (2007), we have derived the periods and amplitudes of the rotational modulation of brightness and color for 31 CTTS; for six of them, the periods have been revealed for the first time. The inclinations of the rotation axis and equatorial rotational velocities of CTTS have been determined. We show that the known periods of brightness variations for some of the CTTS are not the axial rotation periods but are the Keplerian periods near the inner boundary of the dusty disk. We have found that the angular velocity of CTTS with a mass of 0.3--3 $M_\odot$ in the Taurus--Auriga complex remains constant in the age range 1--10 Myr. CTTS on radiative evolutionary tracks rotate faster than completely convective CTTS. The specific angular momentum of CTTS depends on the absolute luminosity in the H$\alpha$ line.

\subsection*{1. Introduction}

One of the fundamental parameters of a star is its angular velocity. At early pre-main-sequence evolutionary stages, mass accretion and gravitational contraction of a star should cause it to spin up due to the conservation of angular momentum. In this case, a solar-mass star with a typical accretion rate of 10$^{-7}$ $M_\odot$ yr$^{-1}$ would reach a rotational velocity close to the critical one by an age of about 1 Myr (Hartmann and Stauffer 1989). However, it is well known from observations of T Tauri stars (TTS) in young clusters that their rotational velocities account for only 10--20\% of the critical ones. Obviously, there should exist excess angular momentum removal mechanisms. At present, two possible mechanisms are considered: (1) mass loss by a star in the form of a magnetized stellar wind carrying away part of the angular momentum (see, e.g., Ferreira et al. 2000; Matt and Pudritz 2005); and (2) the interaction of the stellar magnetic field with an ionized gas in an accretion disk. The magnetic field couples the star and the disk regions rotating with a Keplerian angular velocity, which leads to stellar rotation regulation over 1--10 Myr, as long as disk accretion continues (Konigl 1991; Collier Cameron and Campbel 1993; Shu et al. 1994; Bouvier et al. 1997). This model, commonly called ''disk locking'', is confirmed by the inverse relationship between stellar angular velocity and near-infrared excess, suggesting the presence of a disk. In addition, the rotational velocity of a star can also change due to a change in its moment of inertia, which depends on the internal structure of the star. 

The rotation periods have been well studied for TTS that are not coupled to their accretion disks. These are the so-called weak-line TTS (WTTS). Cool spots on the surface of such stars (an analog of sunspots) can occupy a significant area. This leads to pronounced modulation of their brightness and allows the rotation period to be determined from photometric series of observations. However, the stars where mass accretion still continues, where the star is coupled to its accretion disk, are of special interest. These are classical TTS (CTTS) with strong emission lines in their spectra. There are both mass accretion and outflow in these stars. Both these processes affect the evolution of the stellar angular momentum. Studies of the rotational velocity in groups of young stars with different ages, from 1 to 100 Myr, show that a young star evolves at a nearly constant angular velocity during the first $\sim$4 Myr (Rebull et al. 2004). 

It is not always possible to determine the rotation period of CTTS from photometric observations, because the photometric variability of CTTS is predominantly irregular in pattern. Magnetospheric accretion gives rise to hot spots on the stellar surface, which can cause rotational modulation of brightness, but only for the short lifetime of such a spot, typically no more than two stellar revolutions (Herbst et al. 2007). The stochastic nature of accretion washes out this periodicity. In addition, obscuration of the star by dust clouds near the inner accretion disk boundary causes photometric variability with approximately the same characteristic time as its rotation. Finally, strong emission lines contribute noticeably to the $U$ and $B$ photometric bands. Therefore, nonstationary processes in the stellar magnetosphere can lead to brightness and color variability. 

However, if a hot spot is located at the pole of a magnetic dipole shifted relative to the rotation pole, then more stable modulation of brightness caused by a change in the visibility of the magnetic pole as the star rotates is possible. Since accretion is not constant in time, such a spot can appear and disappear in different observing seasons, but the brightness oscillations will be coherent: the period and phase will be conserved. This makes it possible to determine the rotation period using large data sets accumulated over 20 or more years. The catalogues by Herbst et al. (1994) and Grankin et al. (2007) are the two best-known photometric catalogues containing data on the variability of young stars. The main difference between these catalogues is that the catalogue by Herbst et al. (1994) contains compiled data, while the catalogue by Grankin et al. (2007) is based on homogeneous observations obtained only at one Maidanak Observatory in Uzbekistan. 

There exist a large number of papers devoted to determining the axial rotation periods of WTTS, CTTS, and Herbig Ae stars located in groups of young stars with different ages (see the review by Herbst et al. (2007)). Recently, Percy et al. (2010) searched for a periodicity in 162 TTS and Herbig Ae stars based on data from the photometric catalogue by Herbst et al. (1994) and determined the rotation periods for many stars, including thirty CTTS. Present day CCD observations of compact groups of young stars, such as the cluster NGC 2264 or the Orion Nebula cluster (ONC), allow the periods of many hundreds of stars to be determined at once. Nearer and, accordingly, more open groups (associations), such as those in the Taurus--Auriga complex, are investigated by the traditional method of photometry of individual stars, which takes a much longer time. On the other hand, these are brighter stars and their physical parameters (temperatures, luminosities, and rotational velocities) have already been studied well enough.

Previously (Artemenko et al. 2010), we investigated the characteristic time of photometric variability for CTTS in the range from 20 to 200 days. This definitely exceeds the axial rotation periods and belongs to the Keplerian rotation periods of inhomogeneities in the accretion disk. Using the method of power-spectrum colorimetry, we found the characteristic variability time to depend on the total luminosity of the ''star + disk'' system: the higher the luminosity, the slower the brightness variations, with the characteristic time being consistent with the Keplerian period at the inner boundary of the dusty disk in this system. Fairly stable Keplerian periods that manifested themselves in several observing seasons were found in the brightness variations of several CTTS. 

Here, we investigate the {\it axial} rotation periods of CTTS. First, we briefly discuss the effects of the rotational modulation of spectroscopic and photometric parameters. Then, we analyze the photometric series of CTTS presented in the catalogue by Grankin et al. (2007) to confirm the known axial rotation periods and to detect new ones in the range from 2 to 20 days. Based on the available and newly obtained data, we reach conclusions about the evolution of the angular momentum of CTTS in the Taurus--Auriga complex.

\subsection*{2. Determining the rotation periods of CTTS}

Here, we use the traditional method of determining the rotation periods from the modulation of brightness and color induced by inhomogeneities (spots) on the stellar surface. Attention should also be given to the rotational modulation of {\it spectroscopic} parameters, which may also turn out to be a decisive argument in some cases. If there are regions of enhanced chromospheric emission on the stellar surface, then periodic variations in the radial velocity of narrow emission lines are observed. Moreover, the effect of antiphase variations in the radial velocity of a star determined from photospheric lines and in the radial velocity of narrow chromospheric emission lines is observed. This is because a narrower component of the local chromospheric emission is present in the profiles of rotationally broadened photospheric lines. At low $V_{\rm{eq}}\sin i$ or at an insufficiently high spectral resolution, this appears as an asymmetry of the photospheric line profile and leads to an apparent variation in stellar radial velocity within no more than $\pm V_{\rm{eq}}\sin i$. This effect was first detected in RW~Aur~A (Petrov et al. 2001a) and then in several more CTTS: DR~Tau, S~CrA~SE, and DI~Cep (Petrov et al. 2011). The presence of a {\it cool} spot also leads to variability of the stellar radial velocity determined from photospheric lines, but this effect is smaller by an order of magnitude (Makarov et al. 2009). Antiphase variations in the radial velocities of photospheric and chromospheric lines can serve as a good method of determining the stellar rotation period, because stellar rotation is obviously the only cause of this effect, while the photometric variability can be caused by many factors.

The most homogeneous and longest series of photometric observations for young stars were obtained at the Maidanak Observatory as part of the ROTOR program (Shevchenko 1989). This long-term program of observations was finished in 2006. As a result, more than 100 thousand observations were accumulated for 379 objects. In particular, data are available for 72 CTTS and 48 WTTS. The final results of CTTS and WTTS observations were published in two papers by Grankin et al. (2007, 2008). However, the phenomenon of rotational modulation was searched for and analyzed only for a sample of WTTS (Grankin et al. 2008). Subsequently, Artemenko et al. (2010) analyzed a sample of CTTS in an effort to find the periods in the range from 20 to 200 days related to the Keplerian rotation near the inner boundary of the dusty accretion disk. 

In this paper, we use the catalogue by Grankin et al. (2007) to analyze the variability of brightness and color for CTTS in the range of periods from 2 to 20 days. Here, we give only the periods that are repeated in more than two observing seasons. A description of the observing conditions and the photometric data are given in Grankin et al. (2007) and are accessible in electronic form at CDS, Strasbourg (anonymous ftp to cdsarc.u-strasbg.fr). All of the data were reduced to Johnson's $UBVR$ system. For stars brighter than $V=12^m$, the rms error of a single measurement is $0.^m$01 in $B,V$ and $R$ and $0.^m$05 in $U$. From this database we selected only the CTTS whose photometric series consisted of series of at least three (a maximum of 23) observing seasons with a season duration from three to six months. Each object was usually observed once every clear night during the season.

The light curves of these stars exhibit both rapid night-to-night variations and slow season-to-season variations. Since we are interested in the possible periods on a time scale from 2 to 20 days, any other longer periodic or cyclic light variations are considered as a low-frequency trend or noise. Apart from the weather conditions, the lunar phases and the object visibility seasons are responsible for the gaps in observations. To remove the low-frequency trend, the seasonal series of observations were fitted by a second-order polynomial and the polynomial values were subtracted from the observed ones.

The data filtered in this way were analyzed by various spectral estimation methods: CLEAN (Roberts et al. 1987), $\chi^2$ (Horne et al. 1986), SCARGLE (Horne and Baliunas 1986), and CORRPSD (Marple, Jr., 1990). Since the power spectra calculated by the methods listed above have similar statistical characteristics, we will discuss the results of our spectral analysis using the classical correlogram CORRPSD method as an example. Figure 1 shows examples of the power spectra and phase diagrams for three CTTS: AA~Tau, LX~Ori, and LkHa~274. The 95\% confidence level relative to the mean level of the power spectrum is denoted by the dashed line. The mean level of the spectrum was determined as a highly smoothed spectral estimate corresponding to 20--40 degrees of freedom. A detailed description of the technique for calculating the confidence intervals can be found in the book by Jenkins and Watts (1971).

\begin{figure}[h]
\epsfxsize=14cm
\vspace{0.6cm}
\hspace{1cm}\epsffile{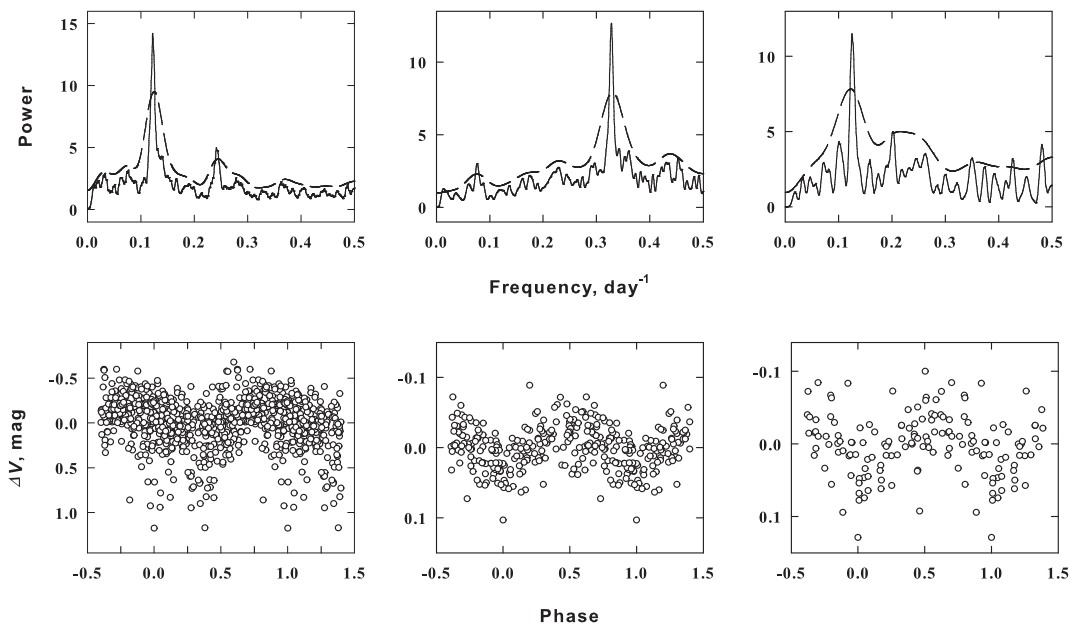}
\caption{\rm \footnotesize { Power spectra and phase diagrams for CTTS. The periods are given in the table. The dashed curve denotes the 95\% probability that the power peaks above this level are non-random.}}
\end{figure}


\begin{table*}
\caption{Stellar parameters, rotation periods, and ratios of the color and brightness amplitudes}
\centering
\scriptsize  
\tabcolsep=2.0mm
\vspace{5mm}\begin{tabular}{|l|c|c|c|c|c|c|c|c|c|c|c|c|} 
\hline
Object & SpT & log $L_*$ & Ref. \dag& $M_*$  & $R_*$ & $V_{eq}\sin i$ & Ref.& $P_{our}$ &  $P_{lit}$ & Ref. & $\sin i$ \ddag & d(B-V)/dV \\
       &     & $L_\odot$ &     & $M_\odot$ & $R_\odot$ & km/s  &     & days     &  days     &     &       & \\
\hline                           
AA Tau    & K7  &  -0.097 & 1  & 0.76 & 1.76 & 12.8 & 15 &  8.19 &  8.20 & 22 & 1.16 & 0.23 $\pm$ 0.13\\
BM And    & K5  &   0.740 & 2  & 1.19 & 4.03 & 38.0 & 6  & 18.58 & 11.00 & 22 & 3.41 & 0.15 $\pm$ 0.17\\
BP Tau    & K7  &  -0.022 & 1  & 0.75 & 1.84 & 13.1 & 15 &  8.10 &  7.60 & 1  & 1.13 & 0.94 $\pm$ 0.23\\
CI Tau    & K7  &  -0.060 & 1  & 0.76 & 1.77 & 13.0 & 16 & 16.10 & 14.00 & 22 & 2.30 & -               \\
DD Tau    & M3  &  -0.469 & 1  & 0.35 & 1.52 & 11.5 & 15 &  5.74 &  5.50 & 22 & 0.85 & 0.46 $\pm$ 0.22\\
DF Tau    & M2  &   0.301 & 3  & 0.39 & 3.55 & 20.0 & 3  &  7.18 &  7.20 & 23 & 0.79 & 0.60 $\pm$ 0.14\\
DG Tau A  & K6  &   0.230 & 1  & 0.90 & 2.23 & 21.7 & 1  &  -    &  6.30 & 1  & 1.20 & -               \\
DI Cep    & G8  &   0.708 & 3  & 1.74 & 2.31 & 20.2 & 17 &  -    &  2.09 & 18 & 0.36 & -               \\
DI Tau    & M0  &  -0.004 & 1  & 0.56 & 2.05 & 10.5 & 1  &  7.65 &  7.70 & 1  & 0.76 & -               \\
DK Tau    & K7  &   0.114 & 1  & 0.74 & 2.17 & 17.5 & 15 &  8.18 &  8.18 & 22 & 1.29 & 0.18 $\pm$ 0.08\\
DL Tau    & M0  &  -0.155 & 3  & 0.57 & 1.80 & 19.0 & 15 &  9.35 &  9.30 & 22 & 1.92 & 0.19 $\pm$ 0.19\\
DM Tau    & M1  &  -0.523 & 1  & 0.48 & 1.28 & 4.00 & 15 & 15.20 &  -    & -  & 0.92 & 0.81 $\pm$ 0.17\\
DN Tau    & M0  &   0.000 & 1  & 0.56 & 2.05 & 12.3 & 15 &  6.32 &  6.20 & 22 & 0.74 & 0.45 $\pm$ 0.17\\
DR Tau    & K7  &   0.230 & 3  & 0.74 & 2.46 & 5.00 & 18 &  -    &  5.00 & 22 & 0.20 & -               \\
DS Tau    & K2  &  -0.222 & 3  & 0.99 & 1.02 & <10  & 3  &  8.08 &  7.70 & 24 &<1.54 & 1.04 $\pm$ 0.21\\ 
EZ Ori    & G3  &   0.771 & 4  & 1.62 & 2.25 & 29.0 & 19 &  2.34 &  1.50 & 19 & 0.59 & -                \\
GG Tau A  & K7  &   0.100 & 5  & 0.74 & 2.16 & 11.5 & 15 &  -    & 10.30 & 25 & 1.07 & -                \\
GI Tau    & K7  &   0.000 & 1  & 0.75 & 1.89 & 12.7 & 15 &  7.09 &  7.00 & 22 & 0.93 & 0.36 $\pm$ 0.16\\
GK Tau    & K7  &   0.146 & 1  & 0.73 & 2.34 & 18.7 & 1  &  4.61 &  4.60 & 22 & 0.72 & -                \\
GM Aur    & K7  &  -0.155 & 3  & 0.77 & 1.56 & 14.8 & 15 &  6.02 &  6.10 & 22 & 1.11 & 0.40 $\pm$ 0.15\\
GW Ori    & G0  &   1.791 & 4  & 3.48 & 6.82 & 40.0 & 3  &  -    &  3.30 & 24 & 0.38 & -                \\
LkHa 191  & K0  &   1.080 & 6,12  & 2.68 & 3.97 & 24.0 & 6  &  2.08 &  -    & -  & 0.25 & -                \\
LX Ori    & K3  &   0.660 & 7  & 1.90 & 2.94 & 29.0 & 7  &  3.05 &  -    & -  & 0.59 & -                \\
RW Aur    & K2  &   0.230 & 8  & 1.48 & 1.71 & 17.2 & 21 &  2.64* & 5.30 & 20 & 1.04 & 0.40 $\pm$ 0.29\\
RY Tau    & K1  &   0.881 & 1  & 2.46 & 3.39 & 48.0 & 16 &  -    &  5.60 & 1  & 1.54 & -                \\
SU Aur    & G2  &   0.799 & 9  & 1.65 & 2.32 & 66.2 & 9  &  -    &  1.76 & 9  & 0.98 & -                \\
T Tau     & K0  &   0.950 & 1  & 2.41 & 3.44 & 20.1 & 1  &  2.81 &  3.00 & 22 & 0.32 & -                \\
UX Tau A  & K2  &   0.000 & 3  & 1.21 & 1.33 & 20.0 & 3  &  -    &  2.70 & 26 & 0.79 & -                \\
UY Aur    & K7  &   0.020 & 5  & 0.75 & 1.90 & 23.8 & 15 &  -    &  6.70 & 24 & 1.64 & -                \\
V1079 Tau & K5  &  -0.131 & 8  & 1.02 & 1.54 & 13.9 & 15 &  6.03 &  5.85 & 27 & 1.06 & 0.32 $\pm$ 0.19\\
V1121 Oph & K5  &   0.176 & 10 & 1.11 & 2.07 & 10.0 & 3  &  8.60 &  8.43 & 22 & 0.81 & -                \\
V2251 Oph & K5  &   0.180 & 11 & 1.11 & 2.07 & -    &    & 14.15 & 14.10 & 22 & -    & 0.49 $\pm$ 0.26\\
V2252 Oph & K3  &   -     & 6  & -    &  -   & 15.0 & 6  &  8.73 &  8.50 & 22 & -    & -                \\
V2503 Oph & K6  &   0.700 & 11 & 0.95 & 3.95 & -    &    &  -    &  3.46 & 28 & -    & -                \\
V360 Ori  & K4  &   0.650 & 7  & 1.62 & 3.20 & 20.0 & 7  &  5.55 &  5.50 & 22 & 0.68 & 0.49 $\pm$ 0.16\\
V625 Ori  & K6  &   0.830 & 12 & 1.00 & 4.67 & -    &    &  5.54 &  -    &    & -    & 0.52 $\pm$ 0.25\\
V649 Ori  & K4  &   0.910 & 12 & 1.80 & 4.30 & -    &    &  4.95 &  5.10 & 22 & -    & -                \\
V687 Mon  & K4  &   1.810 & 12 & 3.60 & 12.0 & -    &    &  7.98 &  -    & -  & -    & -                \\
V866 Sco  & K5  &   0.602 & 10 & 1.14 & 3.34 & 10.0 & 3  &  6.78 &  -    & -  & 0.40 & 0.28 $\pm$ 0.17\\
V895 Sco  & K5  &   0.050 & 11 & 1.15 & 1.98 & -    &    &  3.78 &  3.40 & 22 & -    & -                \\
XZ Tau    & M2  &  -0.770 & 1  & 0.38 & 1.08 & 15.0 & 15 &  3.24 &  2.60 & 1  & 0.89  & -                \\
\hline\hline                                                                                           
CV Cha    & G8  &   0.903 & 3  & 2.04 & 2.94 & 32.0 & 3  &  -    &  4.40 & 3  & 0.93 & -                \\
CW Tau    & K3  &   0.041 & 1  & 1.29 & 1.96 & 27.4 & 1  &  -    &  8.25 & 1  & 2.24 & -                \\
CY Tau    & M1.5&  -0.301 & 1  & 0.43 & 1.65 & 10.0 & 1  &  -    &  7.50 & 1  & 0.88 & -                \\
DE Tau    & M2  &  -0.046 & 3  & 0.39 & 2.33 & 9.70 & 15 &  -    &  7.60 & 25 & 0.62 & -                \\
DH Tau    & M3  &  -0.252 & 1  & 0.30 & 1.94 & 10.0 & 1  &  -    &  7.00 & 22 & 0.70 & -                \\
DO Tau E  & M0  &  -0.102 & 13 & 0.56 & 1.90 & 11.1 & 21 &  -    &  12.5 & 24 & 1.42 & -                \\
HP Tau    & G0  &   1.188 & 1  & 2.12 & 3.42 & 15.4 & 1  &  -    &  5.90 & 1  & 0.52 & -                \\
IQ Tau    & M0.5&  -0.056 & 1  & 0.52 & 2.11 & 11.5 & 1  &  -    &  6.25 & 1  & 0.66 & -                \\
RY Lup    & K4  &   0.415 & 3  & 1.54 & 2.59 & 38.0 & 3  &  -    &  3.80 & 3  & 1.09 & -                \\
SR 9      & K6  &   0.544 & 3  & 0.92 & 3.24 & 24.0 & 3  &  -    &  6.40 & 3  & 0.92 & -                \\
TW Hya    & K7  &  -0.460 & 14 & 0.81 & 1.22 & 4.00 & 14 &  -    &  3.56 & 14 & 0.23 & -                \\
\hline 
\multicolumn{13}{l}{ }\\
\multicolumn{13}{l}{Note. 1 - Guedel et al. (2007); 2 - Eisner et al. (2007); 3 - Johns-Krull et al. (2000);}\\
\multicolumn{13}{l}{4 - Calvet et al. (2004); 5 - Hartmann et al. (1998); 6 - Padgett (1996); 7 - Wolff et al. (2004);}\\
\multicolumn{13}{l}{8 - Akeson et al. (2005); 9 - DeWarf et al. (2003); 10 - Andrews et al. (2010); 11 - Magazzu et al. (1992);}\\
\multicolumn{13}{l}{12 - Cohen and Kuhi (1979); 13 - Ricci et al. (2010); 14 - Donati et al. (2011); 15 - Cuong Nguyen et al. (2012);}\\
\multicolumn{13}{l}{16 - Cuong Nguyen et al. (2009); 17 - Gameiro et al. (2006); 18 - Petrov et al. (2011); 19 - Gagne et al. (1995); }\\
\multicolumn{13}{l}{20 - Dodin et al. (2012); 21 - Hartmann and Stauffer (1989); 22 - Percy et al. (2010); 23 - Chelli et al. (1999);}\\
\multicolumn{13}{l}{24 - Osterloh et al. (1996); 25 - Johns-Krull (2007); 26 - Bouvier and Bertout (1989); 27 - Bouvier et al. (1995);}\\
\multicolumn{13}{l}{28 - Berdnikov et al. (1991); 29 - Petrov et al. (2001b).}\\
\multicolumn{13}{l}{* - 1/2 of the axial rotation period (see the text).}\\
\multicolumn{13}{l}{\dag\, - References to the spectral type and luminosity of the star.} \\
\multicolumn{13}{l}{\ddag\, - The formal value of sin i calculated from Eq.(1).}

\end{tabular}
\end{table*}

The table gives basic parameters of the CTTS being investigated and their rotation periods, both the previously known ones ($P_{\rm{lit}}$) and those we found ($P_{\rm{our}}$) in this paper. The spectral types and luminosities of the stars were taken from literature, while their masses and radii were determined from the models by Siess et al. (2000). We confirmed the known periods for some of the stars, found slightly different periods for other stars, and revealed the periods for the first time for six stars (AS~205, DM~Tau, LkHa~191, LkHa~274, LX~Ori, V625~Ori). The stars from the catalogue by Grankin et al. (2007) that we investigated are contained in the upper part of the table; the parameters and rotation periods of other CTTS collected from published sources are given in the lower part.

\subsection*{3. Discussion} 

As can be seen from the table, the overwhelming majority of CTTS have periods no longer than 10 days. Several remarks should be made about the periods of individual stars. For RW~Aur~A, the 2.77-day period was determined from the antiphase radial velocity variations of emission and absorption lines (Petrov et al. 2001a).The 2.64-day period close to this value is also revealed by longer series of photometric observations, with the periodicity being present only in $U$ and $B$ and being caused by emission line intensity variations (Petrov et al. 2001b). This period can result from the presence of two chromospheric spots on the stellar surface, probably near the magnetic poles shifted relative to the rotation axis. As the star rotates, one and the other spot alternately appear on the hemisphere facing the observer. Consequently, the axial rotation period of RW~Aur~A is about 5.3 days. Spectropolarimetric observations of RW~Aur (Dodin et al. 2012) revealed a period of 5.6 days. Here, we took the rotation period of RW~Aur to be 5.3 days, because it is based on a longer time interval of observations.

For DI~Cep, a period of about 9 days was previously detected in the H$\alpha$ variations, which was interpreted as the presence of a hot spot in the accretion disk (Ismailov 2004). A shorter period of 2.092 days is detected in the antiphase radial velocity variations (Petrov et al. 2011) and is obviously the axial rotation period of the star. $P$ = 2.092 days and $V_{\rm{eq}}\sin i$ = 20.2 km~s$^{-1}$ at the stellar radius $R_*$ = 2.31 $R_\odot$ (see the table) lead to an inclination $i\,\approx\,20\,^\circ$, i.e., the star is seen pole-on and its equatorial velocity is $\approx$ 57 km~s$^{-1}$. In this case, we always see only one pole and the spectral line shifts are caused by the motion of the hot spot around the rotation pole. A similar case of a star seen pole-on was considered and modeled by Petrov et al. (2011) for DR~Tau, where the effect of antiphase radial velocity variations is clearly present. 

As has been pointed out above, the periodic brightness variations can pertain not only to the axial rotation of a star but also to the rotation of an inhomogeneous gas-dust disk. To distinguish the axial rotation periods of a star and the periods induced by Keplerian rotation in the disk, the inclination of the rotation axis $i$ should be determined. If the period $P$ is in days, the projected equatorial rotational velocity $V_{\rm{eq}}\sin i$ is in km~s$^{-1}$, and the stellar radius $R_*$ is in solar radii, then 

\hspace{5cm}      $\sin i = 0.0195\,P\,V_{\rm{eq}}\sin i/R_*$  \hspace{5cm}   (1)
\vspace{0.3cm}

The angular velocity of a star is equal to the Keplerian velocity at the corotation radius $r_{\rm{co}}\,$, defined by the relation $r_{\rm{co}}^3 = G M_* P^2 / 4\pi^2 $. The corotation radius for CTTS is, on average, several stellar radii. According to the ''disk locking'' model, the interaction of the stellar magnetic field with the ionized gas in the disk causes the stellar rotation to become synchronized with the Keplerian rotation at the distance where the stellar magnetic field is still capable of controlling the motion of the ionized gas in the disk, i.e., the corotation radius and the magnetospheric radius become approximately equal in the course of evolution. The time it takes to reach such an equilibrium regime of rotation is about 1 Myr (Matt et al. 2010).

If the modulation of brightness is caused by effects on the stellar surface or in the magnetosphere rotating with the same angular velocity as the star, then Eq. (1) gives $\sin i$ $\leq$ 1. However, if the modulation is caused by effects in the disk (obscuration by dust clouds), then the formal application of Eq. (1) will give $\sin i > 1$. The inner radius of the dusty disk of CTTS $R_{\rm{in}}$ determined from interferometric measurements is $\sim$0.1\,AU (Millan-Gabet et al. 2007), i.e., about ten radii of a typical CTTS. The effects of obscuration by dust clouds can pertain only to distances $\geq$ $R_{\rm{in}}$. Thus, if $\sin i > 1$ was obtained from Eq. (1), this by no means implies that the period was determined erroneously. This can be an erroneous {\it interpretation} of the period, especially since both axial and Keplerian rotation periods are detected for some TTS (Artemenko et al. 2010).

It should be taken into account that determining the inclination from Eq. (1) can be plagued by significant errors related to the uncertainties in each of the parameters appearing in this formula. The period $P$ and $V_{\rm{eq}}\sin i$ are determined directly from observations. The periods of some CTTS, for example, T Tau, are known with an accuracy of the order of one percent, but the relative accuracy of the period is not so high for most of the stars. As a rule, the errors in the period are due to the limited extent and sampling of the series of observations. For the TTS for which the periods were determined repeatedly by different authors, the scatter of values is several percent, with a standard deviation of 3--4\%. A maximum scatter of periods ($\pm$13\%) was observed for DF~Tau.

The radii of CTTS are determined indirectly, from their temperatures and luminosities. Since the effective temperature $T_{\rm{eff}}$ and bolometric luminosity $L_{\rm{bol}}$ of a star are the main parameters when comparing with the models of stars, special attention has always been given to analyzing the errors in these quantities. Here, we refer to the most recent studies in this field. When determining the inclination errors, we are interested primarily in the errors in the stellar radius and $V_{\rm{eq}}\,\sin i$.

Hartmann (2001) discussed in detail the errors in the TTS temperature, luminosity, and radius and their influence on the age determination. In this case, the uncertainties in the distance and interstellar extinction, the physical variability of stars, the possible presence of a secondary component, spots on the surface, and the spectrum veiling are taken into account. It was shown that, typically, the relative error of the bolometric luminosity is $\delta$\,lg\,$L_{\rm{bol}} = 0.16$ and the relative error of the radius is $\delta$\,lg\,$R$ = 0.08 (i.e., $\delta R$ = 20\%). A similar discussion of the errors in the basic parameters of TTS is given in Rebull et al. (2002), where the authors provide typical relative errors: $\delta$\,lg $L_{\rm{bol}} = 0.08$ and $\delta$\,lg $R$ = 0.05 ($\delta R$ = 12\%) for TTS in the range of spectral types K5--M2. The values of $V_{\rm{eq}}\,\sin i$ and their errors for a large number of TTS were recently determined by Cuong Nguyen et al. (2012): for a sample of classical TTS, the mean relative error is $\delta$\,lg\,$V_{\rm{eq}}\,\sin i$ = 0.03 ($\delta V_{\rm{eq}}\,\sin i$ = 7\%).


In our paper, we adopted the following relative errors: $\delta R = 20\,\%$, $\delta P = 10\%$, and $\delta V_{\rm{eq}}\sin i = 10\%$. Given these errors, the relative error is $\delta \sin i\approx$ 24\,\%. This is a rather large error. Nevertheless, large inclinations ($i>  60 \deg$) can be distinguished from small ones ($i < 30\deg$) with a 90\,\% probability.

\begin{figure}[h]
\epsfxsize=7cm
\vspace{0.6cm}
\hspace{5cm}\epsffile{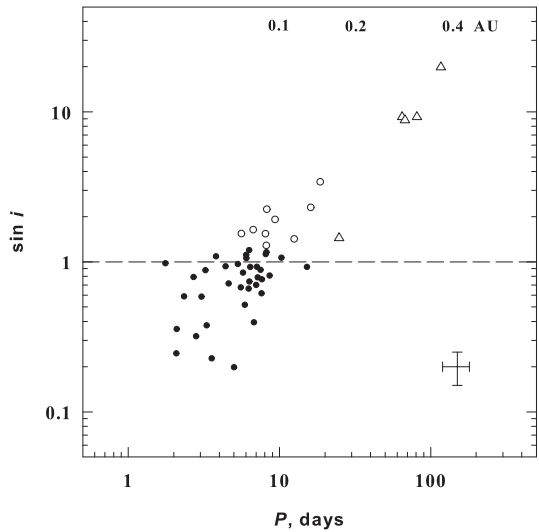}
\caption{\rm \footnotesize {Inclination of the stellar rotation axis versus period. The filled circles, open circles, and triangles indicate, respectively, the axial rotation periods, Keplerian periods, and Keplerian periods found by Artemenko et al. (2010). Typical errors along both axes are indicated in the lower right corner. Approximate semi-major axes of a Keplerian orbit are given at the top.}}
\end{figure}

In Fig. 2, $\sin i$ is plotted against the period $P$. The bar sizes in the lower right corner correspond to the above errors. One might expect the axial rotation periods and Keplerian periods to form two different groups, but it turned out that there is one common sequence: as the period increased, $\sin i$ systematically grows. Most of the data fall into the region of possible axial rotation periods ($\sin i$ $\leq$ 1). 

Strictly speaking, the group of points in the region $\sin i = 1 \pm 1\sigma$ can be attributed with an equal probability to both Keplerian and axial rotation periods. However, given that there should be more axial rotation periods, because they are much easier to detect, we can draw an arbitrary boundary at the level $\sin i = 1 + 1\sigma$ above which the periods are Keplerian with a higher probability. About ten stars fall into this region.

As has been pointed out above, both a temperature inhomogeneity on the stellar surface (a cool or hot spot) and obscuration by dust clouds near the inner boundary of the accretion disk can be responsible for the periodic variability of brightness and color. These sources of variability can be distinguished by the amplitudes of the periodicity in the brightness and color variations. In the case of obscuration by dust clouds, the ratio of the amplitudes of the $B-V$ color and the $V$ magnitude is  $d(B-V)/dV$ $\approx$ 0.28 (Gahm et al. 1989). Smaller and larger $d(B-V)/dV$ correspond to cool and hot spots, respectively (Bouvier et al. 1995). The last column of the table gives the ratios of the amplitudes $d(B-V)/dV$ of the periodic signal for those stars where the period is seen in both brightness and color variations. For the {\it axial} rotation periods (s$\sin i$ $\leq$ 1), the ratio $d(B-V)/dV$ is within the range from 0.28 to 0.81, on average, $d(B-V)/dV = 0.50\,\pm\,0.17$. In this case, we observe mainly hot spots on the stellar surfaces. The Keplerian periods (including those found by Artemenko et al. (2010)) are, on average, $d(B-V)/dV = 0.28\,\pm\,0.17$, corresponding to the case of obscuration by dust clouds.

As can be seen from Fig. 2, the axial rotation periods of CTTS lie within the range 2--15 days. The distribution in $\sin i$ corresponds to a random distribution in inclination, except for the smallest inclinations ($\sin i < 0.2$), at which the modulation of brightness is difficult to detect. The continuous transition from axial rotation periods to Keplerian ones suggests that the rotational modulation of CTTS brightness arises not only on the stellar surface but also in the magnetosphere and the accretion disk, starting from its inner boundary and further out. The minimum inner radius of the dusty disk of CTTS where the star can already be obscured by dust clouds is 0.07--0.1 AU (Millan-Gabet et al. 2007). At a mean mass of the investigated CTTS of about one solar mass, the Keplerian period at this radius is 7--11 days, which is consistent with the observed boundary (8 $\pm$ 2 days) between the axial and Keplerian rotation periods in Fig. 2. 

Our analysis allows the evolution of the angular velocity with the stellar age to be traced. In this case, we use only the stars with known $V_{\rm{eq}}\,\sin i$ and inclination $i$. This makes it possible to choose stars with {\it axial} rotation periods and to determine the equatorial velocity $V_{\rm{eq}}$. Most of the stars in the table belong to the Taurus--Auriga complex and have ages in the range 1--10 Myr. In this time interval of evolution toward the main sequence, the radius of a solar-mass star decreases approximately by a factor of 2.2. This range is large enough to determine how the rotational velocity changes with radius as the star evolves.

Let us consider simple relations: the angular velocity of a star $\Omega~=~2\pi/P$, the angular momentum $J~\sim~M_*\,R_*^2\,\Omega$, and the linear equatorial rotational velocity $V_{\rm{eq}}=2 \pi R_*/P$. If a star contracts with the conservation of its angular momentum, then the linear equatorial rotational velocity should increase with decreasing radius: $J=const $,   $V_{\rm{eq}} \sim 1/R_*$. If, alternatively, it contracts with a constant angular velocity, then the linear equatorial rotational velocity should decrease with decreasing stellar radius: $P=const$,  $V_{\rm{eq}}\sim R_*$. 

\begin{figure}[h]
\epsfxsize=7cm
\vspace {0.6cm}
\hspace{5cm}\epsffile{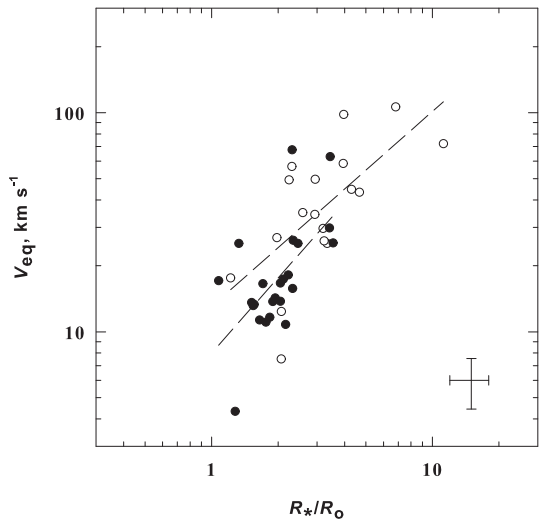}
\caption{\rm \footnotesize {Equatorial rotational velocity versus stellar radius. The filled circles are CTTS in the Taurus--Auriga complex; the open circles are CTTS in Orion and other complexes. The dashed lines are the regression lines.}}
\end{figure}

In Fig. 3, the equatorial rotational velocity is plotted against the stellar radius. Obviously, the rotational velocity $V_{\rm{eq}}$ decreases with decreasing stellar radius. The regression line essentially coincides with what should be in the case of evolution with a constant angular velocity ($P=const$), both in the sample of stars from the Taurus--Auriga complex and in other complexes. The probability of a random correlation is less than 0.5\,\%. This result is consistent with what was obtained from other groups of stars (Rebull et al. 2004). Recall that we consider only CTTS, i.e., the stars that have not yet lost their connection with the accretion disk.

\begin{figure}[ht]
\epsfxsize=15cm
\vspace{0.6cm}
\hspace{0.5cm}\epsffile{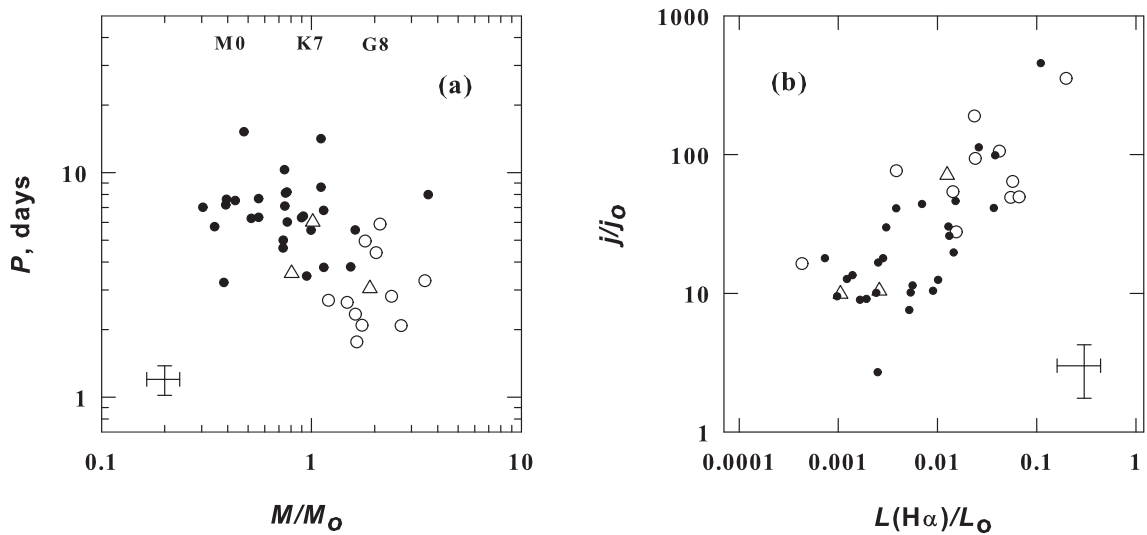}
\caption{\rm \footnotesize (a) Rotation period versus stellar mass. The filled circles are the stars on convective tracks, the open circles are the stars on radiative tracks, and the triangles are the stars at the turning point from the convective track to the radiative one. Approximate spectral types are indicated at the top. (b) Specific angular momentum of a star (in solar units) versus absolute luminosity in the $H_\alpha$ line. The designations are the same.}
\end{figure}

The significant scatter of points in Fig. 3 is caused, in particular, by a difference in stellar masses. Stars of the same age differing in mass lie in different segments of the evolutionary track and have different internal structures. Less massive stars lie on the convective Hayashi track, while more massive stars lie predominantly on the radiative track. In Fig. 4a, the rotation period is plotted against the stellar mass. It can be clearly seen that the more massive stars rotate faster. The probability of a random correlation does not exceed 0.5\,\%. In the group of stars on radiative tracks, the mean rotation period is $P$ = 3.4 $\pm$ 1.5 days, while the completely convective stars have a mean period of 7.2 $\pm$ 2.6 days. According to the ''disk locking model'', the equilibrium rotational velocity depends on the strength of the poloidal magnetic field coupling the star to the disk and on the accretion rate. One of the indicators of accretion is the H$\alpha$ emission. The relations between the accretion luminosity and absolute luminosity in the H$\alpha$ line are given in Fang et al. (2009).

Figure 4b shows the relation between the absolute luminosity in H$\alpha$ and the specific angular momentum of a star $j~=~J/M_*$ in solar units. The error bar on the luminosity axis corresponds to a 30\,\% error in the mean equivalent width of the H$\alpha$ emission line. The probability of a random correlation is less than 0.1\,\%. Stars with a higher luminosity in H$\alpha$ have a larger specific angular momentum, which confirms the role of accretion in the evolution of stellar rotation. 

Another factor is the magnetic field geometry dependent on the internal structure of a star. Spectropolarimetric observations of CTTS showed that the magnetic field configuration in completely convective stars is predominantly dipolar, while the field geometry in stars with a radiative core is more complex (Donati et al. 2008, 2010, 2011). As a result, the large-scale field component in stars with a radiative core is weakened. This leads to the establishment of a smaller radius of the inner accretion disk boundary and to the transfer of angular momentum from the disk to the star (Donati et al. 2011). This mechanism can be responsible for the increase in rotational velocity when CTTS reaches the radiative track.

The masses of our investigated stars lie within the range 0.3--3$M_\odot$. The braking mechanism is known to be most efficient in intermediate-mass TTS, 0.5--1.3$M_\odot$ (Donati et al. 2011). The observations of low-mass TTS ($<\,0.5M_\odot$) in different clusters performed in recent years have shown that the stars rotate faster with decreasing mass (Herbst et al. 2007). Our results show that the stars spin up also in the direction of increasing mass ($>\,1.3M_\odot$), i.e., the dependence of the angular velocity on mass is nonmonotonic.

\subsection*{4. Conclusions}

(1) The rotational modulation of brightness of CTTS is caused by effects not only on the stellar surface (cool and hot spots) but also in the accretion disk (obscuration by dust clouds). 

(2) In the star-forming region in the Taurus--Auriga complex, CTTS in the mass range 0.3--3\,$M_\odot$ retain an approximately constant angular velocity during the first 10 Myr, suggesting the existence of an efficient angular momentum regulation mechanism.
This conclusion is consistent with what is known from present-day studies of the rotation of TTS in the Orion Nebula cluster and other groups of young stars.

(3) The specific angular momentum of CTTS depends on the absolute luminosity in the H$\alpha$ line, indicating a key role of accretion in regulating the angular momentum.

(4) CTTS on radiative tracks rotate, on average, faster than completely convective CTTS. This is probably due to a change in the star's magnetic field configuration when it reaches the radiative track and, as a result, a change in the role of magnetospheric accretion in regulating the stellar rotation.

\newpage

\subsection*{References}

\noindent
1. R. L. Akeson, A. F. Boden, and J. D. Monnier, Astrophys.J. 635, 1173 (2005).

\noindent
2. S. M. Andrews, D. J. Wilner, A. M. Hughes, et al.,Astrophys. J. 723, 1241 (2010).

\noindent
3. S. A. Artemenko, K. N. Grankin, and P. P. Petrov,Astron. Rep. 54, 163 (2010).

\noindent
4. L. N. Berdnikov, K. N. Grankin, A. V. Chernyshev, et al., Astron. Lett. 17, 23 (1991).

\noindent
5. J. Bouvier and C. Bertout, Astron. Astrophys. 211, 99 (1989).

\noindent
6. J. Bouvier, E. Covino, O. Kovo, et al., Astron. Astrophys. 299, 89 (1995).

\noindent
7. J. Bouvier, M. Forestini, and S. Allain, Astron. Astrophys. 326, 1023 (1997).

\noindent
8. N. Calvet, J. Muzerolle, C. Briceno, et al., Astrophys. J. 128, 1294 (2004).

\noindent
9. A. Chelli, L. Carrasco, R. Mujica, et al., Astron. Astrophys. 345, L9 (1999).

\noindent
10. M. Cohen and L. V. Kuhi, Astrophys. J. Suppl. Ser. 41, 743 (1979).

\noindent
11. A. Collier Cameron and C. G. Campbel, Astron. Astrophys. 274, 309 (1993).

\noindent
12. D. Cuong Nguyen, R. Jayawardhana, M. H. van Kerkwijk, et al., Astrophys. J. 695, 1648 (2009).

\noindent
13. D.CuongNguyen, A. Brandeker, M. H. van Kerkwijk, et al., Astrophys. J. 745, 119 (2012).

\noindent
14. A. V. Dodin, S. A. Lamzin, and G. A. Chuntonov, Astron. Lett. 38, 167 (2012).

\noindent
15. J.-F. Donati, M. M. Jardine, S. G. Gregory, et al., Mon. Not. R. Astron. Soc. 386, 1234 (2008).

\noindent
16. J.-F. Donati, M. B. Skelly, J. Bouvier, et al., Mon. Not. R. Astron. Soc. 402, 1426 (2010).

\noindent
17. J.-F. Donati, J. Bouvier, F. M. Walter, et al., Mon. Not. R. Astron. Soc. 412, 2454 (2011).

\noindent
18. J. A. Eisner, L. A. Hillenbrand, R. J. White, et al., Astrophys. J. 669, 1072 (2007).

\noindent
19. M. Fang, R. van Boekel, W. Wang, et al., Astron. Astrophys. 504, 461 (2009).

\noindent
20. M. Fernandez, E. Ortiz, C. Eiroa, et al., Astron. Astrophys. Suppl. Ser. 114, 439 (1995).

\noindent
21. J. Ferreira, G. Pelletier, and S. Stefan, Mon. Not. R. Astron. Soc. 312, 387 (2000).

\noindent
22. M. Gagne, J.-P. Caillault, and J. R. Stauffer, Astron. Astrophys. 445, 280 (1995).

\noindent
23. G. F. Gahm, C. Fischerstrom, K. P. Lindroos, et al., Astron. Astrophys. 211, 115 (1989).

\noindent
24. J. F. Gameiro, D. F.M. Folha, and P. P. Petrov, Astron. Astrophys. 445, 323 (2006).

\noindent
25. K. N. Grankin, S. Yu. Melnikov, J. Bouvier, et al., Astron. Astrophys. 461, 183 (2007).

\noindent
26. K. N. Grankin, J. Bouvier, W. Herbst, et al., Astron. Astrophys. 479, 827 (2008).

\noindent
27. M. Guedel, K. R. Briggs, K. Arzner, et al., Astron. Astrophys. 468, 353 (2007).

\noindent
28. L. Hartmann, Astron. J. 121, 1030 (2001). 

\noindent
29. L.W. Hartmann and J. R. Stauffer, Astron. J. 97, 873 (1989).

\noindent
30. L. Hartmann, N. Calvet, E. Gullbring, et al., Astrophys. J. 495, 385 (1998).

\noindent
31. W. Herbst, D. K. Herbst, E. J. Grossman, et al., Astron. J. 108, 1906 (1994).

\noindent
32. W. Herbst, J. Eisloffel, R.Mundt, et al., in Protostars and Planets V, Ed.by B. Reipurth, D. Jewitt, and K. Keil (Univ. of 
Arizona Press, Tucson, 2007), p. 297.

\noindent
33. J. H. Horne and S. L. Baliunas, Astrophys. J. 302, 757 (1986).

\noindent
34. K. Horne, R. A. Wade, and P. Szkody, Mon. Not. R. Astron. Soc. 219, 791 (1986).

\noindent
35. G. A. J. Hussain, A. Collier Cameron,M.M. Jardine, et al.,Mon. Not. R. Astron. Soc. 398, 189 (2009).

\noindent
36. N. Z. Ismailov, Astron. Rep. 48, 393 (2004).

\noindent
37. G. Jenkins and D. Watts, Spectral Analysis and its Applications (Holden-Day, San Francisco, 1966; Mir, Moscow, 1971), vol. 1.

\noindent
38. C.M. Johns-Krull, Astrophys. J. 664, 975 (2007).

\noindent
39. C. M. Johns-Krull, J. A. Valenti, and J. L. Linsky, Astrophys. J. 539, 815 (2000).

\noindent
40. A. Konigl, Astrophys. J. Lett. 370, L39 (1991).

\noindent
41. A. Magazzu, R. Rebolo, and Y. V. Pavlenko, Astrophys. J. 392, 159 (1992).

\noindent
42. V. V. Makarov, C. A. Beichman, J. H. Catanzarite, et al., Astrophys. J. Lett. 707, L73 (2009).

\noindent
43. S. L. Marple, Jr., Digital Spectral Analysis with Applications (Prentice Hall, Englewood Cliffs, NJ, 1987; Mir, Moscow, 1990).

\noindent
44. S. Matt and R. E. Pudritz, Astrophys. J. 632, L135 (2005).

\noindent
45. S. P.Matt, G. Pinzon, R. de la Reza, et al., Astrophys. J. 714, 989 (2010).

\noindent
46. R. Millan-Gabet, F. Malbet, R. Akeson, et al., in Protostars and Planets V, Ed. by B. Reipurth, D. Jewitt, and K. Keil (Univ. of Arizona Press, Tucson, 2007), p. 539.

\noindent
47. M. Osterloh, E. Thommes, and U. Kania, Astrophys. J. 714, 989 (2010).

\noindent
48. D. L. Padgett, Astrophys. J. 471, 847 (1996).

\noindent
49. J. R. Percy, S.Grynko, and R. Senevirante,Mon. Not. R. Astron. Soc. 122, 753 (2010).

\noindent
50. P. P. Petrov, G. F. Gahm, J. F. Gameiro, et al., Astron. Astrophys. 369, 993 (2001a).

\noindent
51. P. P. Petrov, J. Pelt, and I. Tuominen, Astron. Astrophys. 375, 977 (2001b).

\noindent
52. P. P. Petrov, G. F. Gahm, H. C. Stempels, et al., Astron. Astrophys. 535, 6 (2011).

\noindent
53. L. M. Rebull, S. C. Wolff, S. E. Strom, and R. B. Makidon, Astron. J. 124, 546 (2002).

\noindent
54. L. M. Rebull, S. C. Wolff,S. E. Strom, et al., Astron. J. 127, 1029 (2004).

\noindent
55. L. Ricci, L. Testi, A. Natta, et al., Astron. Astrophys. 512, 15 (2010).

\noindent
56. D. H. Roberts, J. Lehar, and J. W. Dreher, Astron. J. 93, 968 (1987).

\noindent
57. L. E. DeWarf, J. F. Sepinsky, and E. F. Guinan, Astrophys. J. 590, 357 (2003).

\noindent
58. V. S. Shevchenko, Ae/Be~Herbig's Stars (Fan, Tashkent, 1989).

\noindent
59. F. Shu, J. Najita, E.Ostriker, et al., Astrophys. J. 429, 781 (1994).

\noindent
60. L. Siess, E. Dufour, and M. Forestini, Astron. Astrophys. 358, 593 (2000).

\noindent
61. S. C. Wolff, S. E. Strom, and L. A. Hillenbrand, Astrophys. J. 601, 979 (2004).

\end{document}